\documentclass[conference]{IEEEtran}
\usepackage{url}
\usepackage{xcolor}
\usepackage{alltt}
\usepackage{proof}
\usepackage{subcaption}
\usepackage{bigstrut}
\usepackage{xspace}
\usepackage[pdf]{graphviz}
\usepackage{tikz}
\usetikzlibrary{shapes,trees,positioning,arrows}
\usetikzlibrary{shapes.geometric, arrows.meta, positioning, calc}
\usepackage{amsthm}
\usepackage{amsmath}
\usepackage{mathtools}
\usepackage{mathpartir}
\usepackage{listings, listings-rust}
\usepackage{sourcecodepro}
\usepackage[utf8]{inputenc}
\usepackage[T1]{fontenc}
\usepackage{amsfonts}%

\DeclareMathSymbol{\shortminus}{\mathbin}{AMSa}{"39}

\newcommand{\code}[1]{{\small\ttfamily{#1}}}
\newcommand{\w}[1]{\underline{#1}} 

\DeclareUnicodeCharacter{2227}{$\wedge$}
\DeclareUnicodeCharacter{2261}{$\equiv$}
\DeclareUnicodeCharacter{225F}{$\stackrel{?}{=}$}

\newcommand{\ms}[1]{{\color{red}[MS #1 ]}}

\renewcommand{\clap}{\textsc{Clap}\xspace}
\newcommand{\point}[1]{\par\smallskip\noindent\textbf{#1}.}
\newtheorem{definition}{Definition}[section]
\theoremstyle{remark}
\newtheorem{example}{\bf Example}

\lstdefinestyle{mystyle}{
    basicstyle=\ttfamily\footnotesize,
    breakatwhitespace=false,
    breaklines=true,
    captionpos=b,
    keepspaces=true,
    showspaces=false,
    showstringspaces=false,
    showtabs=false,
    tabsize=2,
}
\PassOptionsToPackage{hyphens}{url}\usepackage{hyperref}
\hypersetup{
    colorlinks=true,
    linkcolor=blue,
    filecolor=magenta,
    urlcolor=cyan,
    pdftitle={Clap: a Semantics-preserving Optimizing eDSL for Plonkish Proof Systems},
    pdfpagemode=FullScreen,
    }

\begin{document}
\sloppy

\title{\clap: a Semantic-Preserving Optimizing eDSL for Plonkish Proof Systems}

\author{
  \IEEEauthorblockN{Marco Stronati}
  \IEEEauthorblockA{Matter Labs}
  \and
  \IEEEauthorblockN{Denis Firsov}
  \IEEEauthorblockA{Matter Labs}
  \and
  \IEEEauthorblockN{Antonio Locascio}
  \IEEEauthorblockA{Matter Labs}
  \and
  \IEEEauthorblockN{Benjamin Livshits}
  \IEEEauthorblockA{Matter Labs, Imperial College London}
}

\maketitle

\begin{abstract}
Plonkish is a popular circuit format for developing zero-knowledge proof systems that powers a number of major projects in the blockchain space, responsible for holding billions of dollars and processing millions of transactions per day.
These projects, including zero-knowledge rollups, rely on highly hand-optimized circuits whose correctness comes at the cost of time-consuming testing and auditing.

In this paper, we present \clap, the first Rust eDSL with a proof system agnostic circuit format, facilitating extensibility, automatic optimizations, and formal assurances for the resultant constraint system.
\clap casts the problem of producing Plonkish constraint systems and their witness generators as a semantic-preserving compilation problem.
Soundness and completeness of the transformation guarantees the absence of subtle bugs caused by under- or over-constraining.
Our experimental evaluation shows that its automatic optimizations achieve better performance compared to manual circuit optimization.
The optimizer can also be used to automatically derive custom gates from circuit descriptions.
\end{abstract}

\section{Introduction}
\label{sec:introduction}

PlonK represents a significant milestone in the recent explosion in the design of zero-knowledge proof systems due to its modularity~\cite{gabizon_plonk_2019}.
PlonK's design is subdivided into~1) an arithmetization step,~2) an information-theoretic Interactive Oracle Proof step, and~3) a cryptographic Polynomial Commitment Scheme (PCS) step.
The arithmetization defines a \emph{system of constraints} of a specific form, which is expressive enough to encode an NP statement of interest in a small number of constraints.
The IOP reduces the satisfiability of the constraint system to the problem of checking a number of equations over polynomials.
Finally, the PCS reduces these checks over these potentially very large polynomials to checks over a small number of points.
This kind of modular approach has inspired a large number of variants in recent years~\cite{ben-sasson_scalable_2018, bowe_recursive_2019, setty_spartan_2020, kattis_redshift_2022, chen_hyperplonk_2023}, by swapping new IOPs or PCSs.
The original arithmetization introduced in the PlonK~\cite{gabizon_plonk_2019} paper, has since been extended with the ability define custom gates \cite{gabizon_proposal_2019} and lookup tables~\cite{gabizon_plookup_2020}, and so it is now referred to as Plonkish.
The Plonkish format has gained widespread usage, because of its expressivity and generality, and therefore it is an ideal target representation to take advantage of muliple ``back-end'' proof systems.

\point{Witness generators}
In a proof system, a constraint system is always accompanied by its corresponding \emph{witness generator}.
The witness generator is a program that, given the input to the statement we want to prove, \emph{computes} the corresponding output.
The \emph{witness} to the statement is a concatenation of input, output, and any intermediate value of the computation.
The witness is then used to resolve all the variables in the constraint system and to verify if the system is satisfied.
In other words, the constraint system \emph{checks} the computation performed by the witness generator.

\point{Common vulnerabilities}
Assuming the developer writes a correct witness generator, there are two classes of vulnerabilities that emerge from the discrepancies between what is being computed and what is being checked.
When a constraint system is not satisfied by a witness produced by the witness generator we say it is \emph{over-constrained}.
On the other hand, we say a constraint system is \emph{under-constrained} if it accepts an invalid witness, or a witness that would never be produced by the witness generator.


These classes of bugs are widespread even in critical systems, as pointed out in recent work ~\cite{chaliasos_sok_2024}.

\point{Domain Specific languages}
Before Plonkish, proof systems were largely based on the R1CS format, which is simpler and more rigid.
Writing constraint systems by hand is tedious and error-prone, so several Domain Specific Langauges have emerged over the years to facilitate the task~\cite{eberhardt_zokrates_2018, ozdemir_circ_2022, chin_leo_2021, belles-munoz_circom_2023}, as well as formal verification techniques~\cite{chin_leo_2021, liu_certifying_2023, pailoor_automated_2023, ozdemir_bounded_2023}.

These techniques did not apply as successfully to Plonkish systems, we believe, because of their expressivity.
To the best of our knowledge, the only high level languages for Plonkish are Noir~\cite{aztec_noir_2024} and o1-js~\cite{mina_o1-js_2024}.
However these languages are designed for application developers and they choose to hide most of the complexity (and expressivity) of Plonkish.

\point{Rust Embedded DSLs}
Among expert developers, a far more popular alternative to write Plonkish constraint systems is to use Rust libraries (or shallowly-embedded DSLs), such as Halo2~\cite{zcash_halo2_2022} or Boojum~\cite{matter_labs_boojum_2023}.
These systems allow for full control of the final layout of the constraint system, allowing to exploit peculiar features of a proof system or specific requirements of the application.
However, this control comes at great complexity cost.

\point{Problem 1} The developer is forced to work on \emph{tabulated} constraint systems.
In Plonkish systems, the constraint system is passed to the proof system as a table of variables and constants.
The same constraint system can be tabulated in different ways, which have a significant impact on performance.
We argue that this level of abstraction is too low for developers to write constraint systems that are the same time \emph{safe} and \emph{efficient}.
Furthermore, once constraint systems are written in such a level of detail, they are hard to reuse even for slightly different tabulation choices.

\point{Problem 2} There is a loose connection between constraint systems and witness generators.
The two are written independently, and their correspondence can be hard to verify, especially for optimized code.
A typical dangerous pattern is to allocate cells in the table, which are correctly populated by the witness generator but not properly constrained afterwards, leading to under-constrained bugs.

\point{Limitations of existing formal methods.}
To the best of our knowledge~\cite{certik_verification_2024, soureshjani_automated_2023} are the only existing works that apply formal verification techniques to Plonkish, in particular for the Halo2 system.
These works, however, take what we would call an ex-post-facto approach.
The circuit developer works in a Rust eDSL, until the constraint system and its witness generator are fully developed. The constraint system is then handed over to the proof engineer, who needs to reverse-engineer a specification from it.
In this approach development and verification are disconnected, simple design choices in the development phase or any modification of the circuits can have drastic effects on verification.

\subsection{Motivating Example}
\label{sec:motivation}

The Boojum proof system, and the circuits it defines, power the ZKsync Era Ethereum Rollup which currently holds more than~\$1B in total value locked~(TVL) according to L2BEAT\footnote{\url{https://l2beat.com/scaling/projects/zksync-era}}. For such a mission-critical system, the time that proof production takes is paramount in order to sustain the required transactional throughput; therefore, a significant number of hand-developed circuit optimizations were applied. These optimizations increase the code complexity and the time and effort needed to audit them before they could be deployed in production.

As a motivating example, we look at the Boojum's function \code{split\_36\_bits\_unchecked} used in the SHA2-256 circuit. This function takes a~36-bit value and returns its decomposition into~32 and~4 bit values. Values, which are ultimately represented as Goldilocks field elements, can be abstracted to represent numbers within a range different to that of the field. To do so safely, these values are usually enforced to be within a declared range (range-check) and then given a precise type in Rust. In this case, however, the function name indicates that the input and output of the function are not range-checked. This exception is justified by the fact that range-checks over the same variables are performed by other functions in the same circuit. Avoiding duplicate checks leads to major improvements in the resulting constraint system, at the risk of introducing subtle under-constrained bugs.
This is an example of a complex interaction between Problem~1 and~2 we identified earlier.
Note that these are not theoretical concerns when applied to ZKsync Era circuits: the ZKsync Era circuits were audited in~2023\footnote{Code4rena \href{https://code4rena.com/reports/2023-10-zksync}{audit} of base layer circuits (October 2023).},\footnote{Spearbit's \href{https://github.com/spearbit/portfolio/blob/master/pdfs/Matter-labs-snark-wrapper-Spearbit-Security-Review.pdf}{audit} of the recursive circuits (November 2023).}, which identified~6 high-risk and~22 medium-risk issues.

In fact, during our experimental evaluation, we discovered a critical under-constrained bug in the Boojum implementation of SHA2-256.
The bug was caused by the function \code{split\_36\_bits\_unchecked}, which introduces~8 variables that were not properly constrained later.
This bug resulted from a pattern of collecting unchecked variables to be checked afterwards. This pattern is used twice, but because of a copy-paste mistake, the same collection is checked twice, while one of the collections is never checked. A more detailed overview of this bug is presented in Figure~\ref{fig:range-checks-sha2} in the Appendix\footnote{The bug was disclosed to and was quickly fixed by the developers.}.
To highlight the complexity of missing checks, Section~\ref{sec:experiments} shows further examples where tracking of unchecked variables must be performed across loop iterations.

\subsection{This Work}
Constraint systems are often referred to as \emph{arithmetic circuits}, to provide a useful mental model to developers.
However, this analogy is imprecise as circuit and constraint systems, while similar, are \emph{not} equivalent.
Circuits are functions, and so they deterministically derive an output from an input, while constraint systems are mathematical relations, an unusual model for developers to work with.
We argue that this subtle difference is related to the discrepancies we discussed between witness generators and constraint systems.

\clap proposes a sharp separation between circuits and constraint systems.
Circuits are partial functions which define the witness generators of our system.
They provide a high-level intuitive model for developers and a convenient semantics for verification.
Constraint systems, on the other hand, are a lower-level representation that is suitable for a proof system but should not be exposed to the developer.
\clap proposes a compilation stack where circuits are the source language and constraint systems the target language.

\point{Semantics-preserving compilation}
Thanks to our architecture, we can formally define the properties of soundness and completeness of our compilation step and we show how they correspond respectively to the absence of under- and over-constrained bugs.
These properties guarantee the correspondence between witness generators and constraint system needed to prevent Problem~2 discussed above.
A model of this translation is proved sound and complete in the Adga proof assistant and, importantly, the proof is completely modular with the addition of new custom gates, provided they are sound and complete individually.

\clap is the first system to cast the problem of producing Plonkish constraint systems and their witness generators as a semantic-preserving compilation problem.
Contrary to existing ex-post-facto approaches, \clap offers the first architecture for Plonkish that achieves soundness and completeness by construction.

\point{Safe automatic optimizations}
As discussed above in Problem~1, trying to manually optimize the table representation of a circuit as it is being defined is a form of premature optimization; doing so can lead to critical bugs, such as missing constraints.
It is also very hard to guarantee safety when composing optimized sub-circuits.
\clap exposes only sound and complete gates to the developers, postponing any optimization step until the circuit has been completely defined.
This conservative approach leads to circuits that are initially inefficient, in part because of redundant checks that would not be present in hand-optimized circuits.
However, once the circuit is completed, \clap can apply whole-circuit optimizations, such as removing duplicate checks, with all the context needed to guarantee their safety.

\point{Circuits reuse}
Gates can also be \emph{converted} by the optimizer in order to target different constraint systems. \clap initially produces the same arithmetic gates introduced by the original PlonK paper. However, when targeting the Boojum proof system, that uses three specialized gates for arithmetics, the optimizer performs the translation automatically.
This way, circuits are not prematurely tailored to a specific proof system and can be reused.

\point{Custom gates generation}
Custom gates are a defining feature of Plonkish arithmetization that allows one to trade prover time for verification time. In some proof system designs, custom gates can reach significant complexity, for example to implement a round of the Poseidon hash function in a single constraint. Instead of writing a custom gate from scratch, our inlining optimizer can be used to automatically ``flatten'' a \clap circuit definition into a custom gate of the required degree.
This novel approach avoids multiple implementations of the same logic, thus saving development and review time.

\point{Experimental evaluation}
We validate our thesis that \clap can produce constraint systems from safe circuits at no additional cost, using Boojum's existing circuits as a reference point.
Boojum circuits are used in production by ZKsync Era, are developed and manually optimized by experts, and are reviewed by third party auditors.
We first show several examples where \clap's code quality improves readability, portability, and safety.
We then test the expressivity and efficiency of \clap on an implementation of the Poseidon2 and SHA2-256 hash functions, which are among the most hand-optimized circuits and cover a complete spectrum of the techniques used in other circuits as well.
We examine the size of the resulting constraint systems, showing that even if \clap starts with significantly more constraints, built-in optimization produces~10\% fewer constraints than corresponding hand-optimized circuits.

\subsection{Contributions}
This paper makes the following contributions.
\begin{enumerate}

\item we propose the first semantic-preserving compilation approach for Plonkish called \clap. We capture the absence of over- and under-constrained bugs as soundness and completeness of our compilation.

\item we prove in the Adga proof assistant that our circuits are sound and complete, provided they are built from sound and complete components;

\item we develop five optimizations that reduce circuit size, handle potentially unsafe gates, and convert circuits to target different proof systems;

\item we show how to automate the creation of \emph{custom gates} using a novel optimization we call \emph{flattening}.

\item we test our Rust eDSL implementation on the Poseidon2 and SHA2-256 hash functions, writing safe circuits all the while producing constraint systems~10\% smaller than the hand-optimized Boojum baseline implementations.
In particular, for SHA2-256, we were able to remove~3 unsafe functions, which caused critical bugs in the past, responsible for more than a thousand potentially under-constrained rows.

\end{enumerate}
Both the Agda model and the Rust implementation can be found at
\url{https://github.com/matter-labs/research-public/}.

\subsection{Paper Organization}

Section~\ref{sec:from-programs-to-proofs} introduces the problem of building constraint systems and their corresponding witness generators, and gives an overview of \clap architecture, from eDSL to the proof system.
Section~\ref{sec:source} describes circuits and witness generators.
Section~\ref{sec:cs} defines constraint systems, their satisfiability, and compilation from circuits as well as their tabulation.
Section~\ref{sec:optimizations} describes all optimizations performed by \clap, together with a definition of semantic preservation.
Section~\ref{sec:sem-pre-compilation} defines soundness and completeness of \clap compilation as well as the Agda model.
Section~\ref{sec:experiments} showcases examples where \clap improves safety and readability of Boojum circuits, examines our circuit implementations of Poseidon2 and SHA2-256 as well as our Poseidon2 custom gate.
Section~\ref{sec:related} compares \clap to related work.
Section~\ref{sec:future} mentions some possible future directions.
Finally, Section~\ref{sec:conclusions} concludes.

\section{From Programs to Proofs}
\label{sec:from-programs-to-proofs}

\point{Constraint systems and witness generators}
Typical proof systems involve constraint systems defined on addition and multiplication over a finite field $\mathbb{F}$.
However, it is common to refer to these constraint systems as circuits, which offers a more computational analogy.
Circuits also hint at the fact that in this model computations are finite (e.g. no \code{while} loops are allowed) and also that the cost of each execution is the same, as it corresponds to the size of the circuit (e.g. all branches are always executed).

A constraint system is a system of equations of a specific structure over variables in $\mathbb{N}$ and constants in $\mathbb{F}$.
Variables contain references to positions within a vector of field elements referred to as the \emph{witness}.
The witness contains an assignment to all the variables of the constraint system.
In the circuit interpretation, variables correspond to the wires connecting the gates.
To avoid confusion, we denote with an underscore any name or value in $\mathbb{N}$, e.g. \w{r} or \w{0},  all other numerical values are field elements in $\mathbb{F}$.

In the following we sometimes refer to a witness, or its parts, as \emph{trace}.
This corresponds to the intuition of the trace of execution of a circuit, that starts populated only by inputs and is appended with new intermediate variables as the circuit is executed, including eventually the outputs of the computation.
This process is commonly referred to as \emph{witness generation}. We mention two performance-related observations.
\begin{itemize}
\item \itemsep=0pt
The generation of a constraint system can be computationally intensive.
However, it is performed only once, before the setup phase of the proof system.
The goal is to produce the smallest possible constraint system.
\item
The trace generation step needs to be very efficient, as it is recomputed for every input, and thus it contributes to proving time.
In practice, this contribution can be sizable, so it is important to reduce the trace generation time.
\end{itemize}

\begin{figure}[bt]
  \begin{subfigure}[b]{.5\columnwidth}
\begin{lstlisting}[basicstyle=\footnotesize\ttfamily]
fn chained_add<F:Field>(
  e : &mut Env<F>,
  i1 : Repr<F, SValue>,
  i2 : Repr<F, SValue>,
  i3 : Repr<F, SValue>)
  -> Repr<F, SValue> {
  let m = e.mul(i0, i1);
  e.add(m, i3)
}
\end{lstlisting}
    \caption{Program}
    \label{fig:chained-gates-program}
  \end{subfigure}
  \begin{subfigure}[b]{.4\columnwidth}
    \resizebox{4cm}{!}{%
    \begin{tikzpicture}[
      very thick,
      node distance=1.3cm,
      on grid,
      dang/.style={},
      comp/.style={circle,draw},
      ]
      \node [dang] (o) {};
      \node [comp] (cb) [left=of o] {\code{\tiny arith}}; 
      \node [comp] (ca) [left=of cb,yshift=0.5cm] {\code{\tiny arith}} ; 
      \node [dang] (i0) [left=of ca,yshift=0.5cm] {} ;
      \node [dang] (i1) [below=of i0] {} ;
      \node [dang] (i2) [below=of i1] {} ;

      \path
      (cb) edge[->] node[above]{\w{4}} (o)
      (ca) edge[->] node[above]{\w{3}} (cb)
      (i0) edge[->] node[above]{\w{0}} (ca)
      (i1) edge[->] node[above]{\w{1}} (ca)
      (i2) edge[->] node[above]{\w{2}} (cb)
      ;
    \end{tikzpicture}
    }
    \caption{Circuit}
    \label{fig:chained-gates-circuit}
  \end{subfigure}\\
  \begin{subfigure}[t]{\columnwidth}
    \centering\small
    \[0*t[\w{0}] + 0*t[\w{1}] -1*t[\w{3}] + 1*t[\w{0}]*t[\w{1}] + 0 = 0\]
    \[1*t[\w{2}] + 1*t[\w{3}] -1*t[\w{4}] + 0*t[\w{0}]*t[\w{1}] + 0 = 0\]
    \caption{Constraint system.}
    \label{fig:chained-gates-cs}
  \end{subfigure}\\
  \begin{subfigure}[t]{\columnwidth}
    \centering\small
    \begin{tabular}{c c c | c c c c c}
      l & r & o & ql & qr & qo & qm & qc \\\hline
      \w{0} & \w{1} & \w{3} & 0  & 0  & -1 & 1  & 0  \bigstrut\\
      \w{2} & \w{3} & \w{4} & 1  & 1  & -1 & 0  & 0
    \end{tabular}
    \caption{Table.}
    \label{fig:chained-gates-table}
  \end{subfigure}
  \caption{Example: chained gates.}
  \label{fig:chained-gates}
\end{figure}

\begin{example}[Chained gates]
\label{ex:chained-gates}
As an end-to-end example, let us look at a system supporting only the original PlonK arithmetic gate \code{arith}, which can be encoded as the identity
\begin{align*}
  id\_arith &(l~r~o, ~ql ~qr ~qo ~qm ~qc) \coloneqq \\
      &ql\cdot l + qr\cdot r + qo\cdot o + qm\cdot l \cdot r + qc = 0
\end{align*}
where $l,~r,~o \in \mathbb{F}$ (for left, right, and out) can be interpreted as the value this gate's wires have in the witness, and where $ql,~qr,~qo,~qm,~qc$ can be interpreted as the constants configuring the behavior of the gate.
The programmer defines the program in Figure~\ref{fig:chained-gates-program} that chains a multiplication and an addition.
The program is converted to a circuit \code{chained\_gates} of type \code{circuit}, depicted in Figure~\ref{fig:chained-gates-circuit} where \w{0}, \w{1} and \w{2} are input wires, \w{4} an output wire and \w{3} an intermediate wire connecting the two gates.
The witness generator, given three field elements as inputs, say,~5,7, and 9, can be run as
\code{gen\_trace chained\_gates [5 7 9] = Some [5 7 9 35 44]}.
Positions in the witness correspond to wires; for example, output~44 is in position \w{4},  as expected.
The circuit is converted to the system of constraints depicted in Figure~\ref{fig:chained-gates-cs} with \code{gen\_cs chained\_gates = cs}.
We can then check the satisfiability of the resulting constraint system using \code{sat cs [5 7 9 35 44]}, which replaces the indexes in the constraint system with the values in the witness, obtaining the following equations: \code{5*7=35} and \code{35+9=44}, which are indeed satisfied.
As the last step, we obtain the table for the proof system of Figure~\ref{fig:chained-gates-table} using \code{tabulate cs}.
\end{example}

\begin{figure*}[tb]
  \centering

  \includegraphics[width=\textwidth]{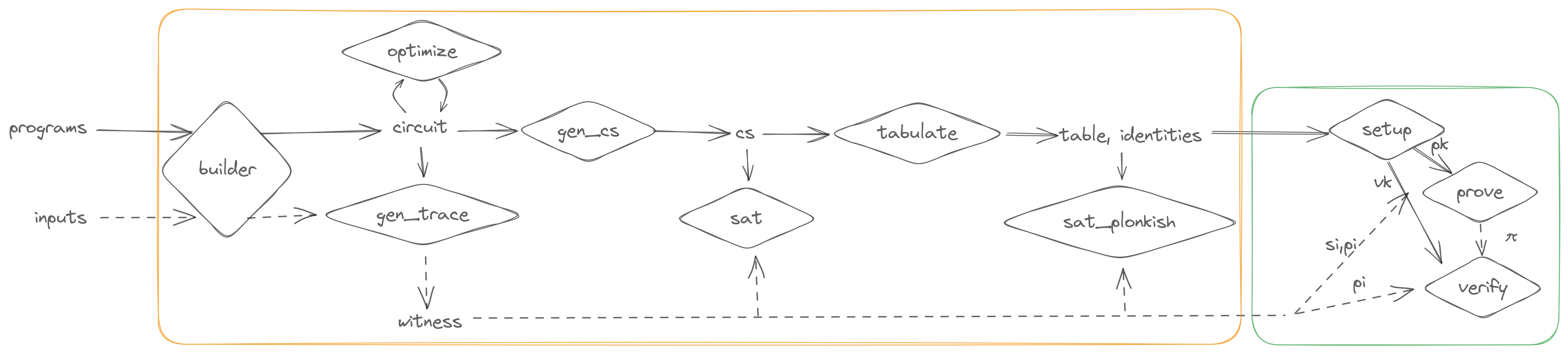}
  \caption{From programs to proofs, \clap in orange, proof system in green.}
  \label{fig:program-to-proof}
\end{figure*}

\point{\clap architecture}
We now describe \clap's compilation architecture depicted in the diagram in Figure~\ref{fig:program-to-proof}.
The developer writes circuits as programs in their main development language, in our case Rust. These programs call into a Rust library we call the \code{builder} which provides a number of \emph{built-in} constructs and combinators to ease the composition of circuits. The builder additionally automatically converts Rust types to field elements to be able to pass inputs to the circuit.

\begin{figure}[ht]
\begin{lstlisting}
gen_trace    : circuit -> trace -> option trace
gen_cs       : circuit -> cs
sat          : cs -> trace -> bool
tabulate     : cs -> list identity * list row
sat_plonkish :
  list identity * list row -> trace -> bool
\end{lstlisting}
\caption{Type signatures for \clap's main functions.}
\label{fig:sem_types}
\end{figure}

The builder constructs a \code{circuit} that encodes, for a given set of gates, their configuration and their wiring.
We give formal semantics to circuits viewing them as trace generators, that is, partial functional programs that receive a trace containing at least their inputs and produce a trace extended with any number of intermediary values and the outputs of the computation.
We define the type \code{trace} of traces and the function \code{gen\_trace}
as the high-level semantics of our circuits.
At the level of circuits, we can also perform optimizations to improve their performance as trace generators or to reduce their size once compiled to constraint systems.

We then define the type \code{cs} that represents constraint systems and give them a formal semantics as the function \code{sat},
that checks if a constraint system is satisfied by a given trace.
Now that we have a source language as the type \code{circuit} and a target language as \code{cs}, we can define a compilation step from the former to the latter as the function \code{gen\_cs}.
Rather than sending a constraint system directly to the proof system, we introduce an intermediary step of \emph{tabulation}, which allows us to have a more flexible definition of \code{cs} that can, in principle, accommodate multiple tabulation strategies.

The original PlonK system supports a single type of constraint on each row, that is, the polynomial identity introduced in the Example~\ref{ex:chained-gates} as $id\_arith$.
However, the Plonkish arithmetization allows to extend a constraint system with custom gates, which correspond to introducing new \emph{sets} of identities.
The tabulation step makes sure that each constraint respects the desired table geometry and builds the set of identities that need to apply over all rows of the table.
The semantics of the proof system is abstractly represented as the function \code{sat\_plonkish}.

Once the program has been transformed into a constraint system, the system is still used for proof generation, as depicted in the dashed edges of the diagram: inputs are serialized by the builder, then run through \code{gen\_trace} to obtain a complete witness, which is split into a public input for the verifier and a public plus private input for the prover.

\point{Gates}
\clap circuits are composed of gates. The set of avaiable gates is defined separately from circuits and each gate consists of:
\begin{itemize}\itemsep=-1pt
\item a sub-case of the \code{gen\_trace} function;
\item a sub-case of the \code{gen\_cs} function;
\item a proof of semantic preservation between \code{gen\_trace} and \code{gen\_cs} as described in Section~\ref{sec:sem-pre-compilation}.
\end{itemize}
Because of Plonkish support for custom gates, we cannot hope to prove semantic preservation once and for all, but we can still provide the necessary scaffolding to build an extensible system.
Notice that the proof engineering work required from the gate designer is limited to their own gate and is therefore completely independent from the rest of the system. As we describe in Section~\ref{sec:agda}, our proof of semantic preservation for the whole system is completely compositional with respect to specific gates.




\section{Circuits and Witness Generators}
\label{sec:source}

\subsection{Circuits and Gates}
\label{sec:circuits}

\clap's circuit format is designed around three main needs.
First, being extensible so that experts can design new gates as in the Plonkish spirit.
Second, the format should be easy to manipulate from a host language so to provide an easy embedding in Rust.
Third, the format should facilitate the proof of semantic preservation presented in Section~\ref{sec:sem-pre-compilation}.
We start by describing the \emph{syntax} of our circuits and their wiring.
To form the reader's intuition, we forget about constraint systems and simply try to capture circuits as we would write them on a whiteboard as in Figure~\ref{fig:chained-gates-circuit}.


\begin{definition}[Gates and circuits]
  We define a \emph{gate} as an identifier together with a tuple of inputs, outputs, and constants $(\vec{\w{i}}, \vec{\w{o}}, \vec{c})$.

  We define a \emph{circuit} as a tree with two kinds of intermediate nodes, sequential and parallel composition, and gates as its leaves.
  In sequential composition, variables defined as outputs in the left subtree can be used as inputs in the right subtree.
  In parallel composition, outputs of the left subtree cannot be used as inputs in the right subtree.
\end{definition}

The constructor \code{seq(c1,c2)} composes sequentially \code{c2} after \code{c1}, which means that \code{c2} can use as input any wire defined in \code{c1}.
Alternatively, parallel composition \code{par(c1,c2)}, imposes that \code{c2} can \emph{not} refer to any wire defined in \code{c1} and vice versa.
This requirement allows to parallelize trace generators, but in the following we refer only to sequential composition for simplicity.

\begin{figure}[ht]
  \begin{subfigure}[b]{0.2\textwidth}
\begin{lstlisting}
circuit ::=
  | nil
  | gate(gate)
  | seq(circuit,
        circuit)
  | par(circuit,
        circuit)
\end{lstlisting}
    \caption{Circuit data type.}
    \label{fig:circuit-definition}
  \end{subfigure}
\begin{subfigure}[b]{0.3\textwidth}
\begin{lstlisting}
gate ::=
  | boolCheck (i) () ()
  | isZero (i) (o) ()
  | arith (l r)
          (o)
          (ql qr qo qm qc)
\end{lstlisting}
\vskip 1em
    \caption{Examples of gates.}
    \label{fig:sample-gates}
  \end{subfigure}
  \caption{Circuits and gates.}
\end{figure}

In Figure~\ref{fig:circuit-definition} we illustrate the definition of a circuit as a tree of gates.
In Figure~\ref{fig:sample-gates} we propose a minimal set of three interesting gates.
This set is much larger in our Rust implementation described in Section~\ref{sec:experiments} or a completely different set could be chosen for a different implementation of \clap.
The \code{arith} gate introduced in \ref{ex:chained-gates}, for instance, requires two inputs, five constants, and provides one output.
The \code{boolCheck} gate is the simplest example of assertion: a gate that returns no output, but can halt the circuit execution.
Lastly, the \code{isZero} gate is the simplest example that makes use of an \emph{oracle}: a gate that manifests auxiliary elements in the trace needed by the constraint system.
We treat oracles in greater detail in Section~\ref{sec:oracles}.


\subsection{Circuit Semantics}

If constraint systems are the right format for proof systems, trace generators are a much more appropriate model for programmers to work with.
Trace generators express the computational nature of a circuit quite well.
Trace generators follow the data flow of wires from inputs to outputs, and populate the trace with the values of all intermediary wires.
Trace generators provide a convenient \emph{functional semantics} for our circuits; note, however, that these functions can be \emph{partial}, not necessarily defined for all possible inputs.

We define a function \code{gen\_trace} that takes the circuit and an initial trace containing at least its inputs and returns a witness for the prover.
The function can fail if the inputs are missing or if an assertion fails during execution, hence the optional result.



\begin{example}[Trace generation]
Let us look at \code{gen\_trace} for the case of the \code{arith} gate.
The values for the two input wires, \w{l} and \w{r}, are retrieved from a trace \code{t}, which we denote with an array-like syntax \code{t[]}.
The output value of the function can be calculated as \code{t[\w{o}] = (ql * t[\w{l}] + qr * t[\w{r}] + qm * t[\w{l}] * t[\w{r}] + qc) / qo} and is written to the trace at position \w{o}.




\end{example}

\subsection{Oracles and Non-Fully-Constrained Traces}
\label{sec:oracles}

In the \code{arith} case seen so far, the witness generator and the constraint system are almost identical.
There are, however, cases when directly computing a value is expensive but checking its correctness is cheap.
For example, computing the inverse of a field element using \code{arith} (and thus only addition and multiplications) is expensive, while its correctness can be easily checked with one multiplication.
In this case \code{gen\_trace} can perform the expensive computation of the inverse and add it to the trace, while the constraint system can remain small because it just needs to check the result.
In these cases we say \code{gen\_trace} acts as an \emph{oracle}.

For example, in the \code{isZero} gate, in the case where the input is not zero, \code{gen\_trace} produces in the trace its inverse.
This way, the constraint system can check with a multiplication that the input could not have been zero because zero has no inverse.
However, if the input is zero, the position in the trace for the inverse is irrelevant and it can contain any value.
This kind of gates are sometimes called non-deterministic, because there are several traces that satisfy their corresponding constraint system.
Notice that the gate is deterministic in its input-output behavior, only the intermediate values can vary, for this reason we prefer the term \emph{non-fully-constrained} traces.

Here are two example evaluations:
\begin{lstlisting}[mathescape]
gen_trace (isZero (i=0) (o=2)) [5] = [5 $\frac{1}{5}$ 0]
gen_trace (isZero (i=0) (o=2)) [0] = [0 0 1]
\end{lstlisting}
Note, however, that another perfectly valid function \code{gen\_trace'} in the second case could return \code{[0 3 1]}, which would still be accepted by the same constraint system.

To sum up, the \code{isZero} gate is a typical example to show:
\begin{itemize}
\item the use of intermediate values: the inverse is not an input or an output of the gate, it is just an auxiliary value needed by the constraint system.
\item non-fully-constrained traces: intermediate values that can take multiple values within deterministic gates.
\end{itemize}
In general, \code{isZero} illustrates the difference between circuits, as deterministic partial functions, and the constraint systems that check them and why this difference can lead to dangerous bugs.

Despite the simplicity of \code{isZero}, the very same problems appear in our motivating example \code{split\_36\_bits\_unchecked}.
This function intuitively returns the decomposition of a~36-bit integer into a~32-bit and a~4-bit value, but in practice the outputs are never computed by the constraint system.
The trace generator computes the outputs and places them in the trace, the constraint system reads them and checks they are consistent with the input.
The danger of oracles is exemplified by systems like Boojum where they can be introduced during circuit definition, making it the responsibility of the developer to guarantee the correspondence with their constraint system.
We propose a more disciplined approach in which, each time a new oracle is needed, a new gate must be defined, together with a proof of semantic preservation between \code{gen\_trace} and \code{gen\_cs}.

Later in Section~\ref{sec:sem-pre-compilation}, we see how non-fully-constrained traces make the definition of soundness slightly more complex, and in turn make the Agda model more challenging.

\section{Constraint Systems and Their Satisfiability}
\label{sec:cs}

In this section, we define the target of our compilation as a constraint systems together with their satisfiability.
We continue the compilation of our example to constraint systems and show how to tabulate it.

In the original PlonK paper~\cite{gabizon_plonk_2019} the arithmetic gate corresponds immediately to a single constraint, that is a polynomial identity over three variables of the specific form already described in Section~\ref{ex:chained-gates}.
We follow the TurboPlonK proposal~\cite{gabizon_proposal_2019} that extended this format to a \emph{conjunction} of polynomial identities of arbitrary degree that share any number of variables and constants.

\begin{definition}[Constrained vector]
  We define a constrained vector \code{cv} as a tuple $(\vec{\w{w}}, \vec{q}, ids)$ of variables, constants, and identities, where each identity is a polynomial function
  $id$ over the values of variables $\vec{\w{w}}$ and $\vec{q}$.
\end{definition}

\begin{definition}[Satisfiability]
  We define function \code{sat\_cv} that given a constrained vector $(\vec{\w{w}}, \vec{q}, ids)$ and a trace $t$:
  \begin{itemize}
  \item resolves all variables in $\vec{\w{w}}$ using the trace $t$
  \item computes $\bigwedge\limits_{i} id_i(t[\vec{\w{w}}],\vec{qs})$
  \end{itemize}
\end{definition}
We define a constraint system \code{cs} simply as a list of constrained vectors and its \code{sat} simply as the conjuction of \code{sat\_cv} over its elements.

\subsection{A Sample of Constrained Vectors}

In order to compile our sample program, we make use of the $id\_arith$ that was defined in Example~\ref{ex:chained-gates}, together with the following boolean identity:
\[
id\_bool (i) \coloneqq i \cdot (i \shortminus 1) = 0
\]

Using these identities, we define three constrained vectors:
\begin{align*}
arith\_cv =(&(\w{l}~\w{r}~\w{o}),~(ql~qr~qo~qm~qc),~id\_arith) \\
boolCheck\_cv =(&(\w{i}),~(),~id\_bool) \\
isZero\_cv =(&(\w{i}~\w{r}~\w{o}),~(0~0~\shortminus\!1~\shortminus\!1~1),\\
           &(id\_arith~id\_bool ))
\end{align*}
Notice that $id\_bool$ could be expressed as $id\_arith,$ but doing so would require a much larger vector.

\subsection{Compilation}

Now that we have two representations, namely \code{circuit} and \code{cs}, and their semantics, respectively \code{gen\_trace} and \code{sat}, we can define a compiler \code{gen\_cs}.
Section~\ref{sec:sem-pre-compilation} shows how this compiler preserves the semantics of our circuits.

Much like \code{gen\_trace}, the behavior of \code{gen\_cs} is completely dependent on the choice of gates that we have in our circuits and of constrained vectors that we have in our constraint systems. For our example, we can use a very simple \code{gen\_cs} that maps each gate to its corresponding constrained vector.

\subsection{Tabulation}
\clap's constrained vectors are independent of each other, which means that they can have different identities, wires, and constants. In a Plonkish proof system, on the other hand, there is a rigid table structure. This table's columns hold either variables or constants, and it's rows are all of the same width. Furthermore, there is a single set of identities that constrains all rows.

A Plonkish proof system can be abstracted as a satisfiability function \code{sat\_plonkish} that takes a list of identities and checks that they all hold within each row of a table.
This function captures the functionality provided by the proof system and it is the lower level of abstraction of our stack.

The last step of the \clap pipeline, \code{tabulate}, turns a constraint system into a corresponding Plonkish table.

\point{Selectors}
A typical way to selectively apply two different identities is to introduce selectors.
A selector is an additional constant that multiplies the entire identity. As the name suggest, this constant is used to select if the identity is applied or not over a given row (setting it to zero ``disables`` the identity for a row).

In our example constraint system, we slightly modify the identities by multiplying them by two new constants \code{qarith} and \code{qbool}.
Introducing new constants, however, increases also the size of the vectors.

In our example tabulation, the two new selectors are both activated for $isZero\_cv$.

\begin{figure*}[tbh]
\centering
\begin{lstlisting}
tabulate arith_cv  = ((\w{l} \w{r} \w{o}), (ql qr qo qm qc 1 0), (qarith * id_arith ; qbool * id_bool ))
tabulate bool_cv   = ((_ _ \w{i}), ( _  _  _  _  _ 0 1), (qarith * id_arith ; qbool * id_bool ))
tabulate isZero_cv = ((\w{i} \w{r} \w{o}), ( 0  0 -1 -1  1 1 1), (qarith * id_arith ; qbool * id_bool ))
\end{lstlisting}
  \caption{Example tabulation.}
  \label{fig:tabulation}
\end{figure*}

Now that all vectors are of the same size and type and share the same set of identities, the constraint system can be passed to the proof system.
There are some tabulation techniques that we leave as future work, such as combining selectors in a tree~\footnote{\url{https://mirprotocol.org/blog/Fast-recursive-arguments-based-on-Plonk-and-Halo}}, repeating constrained vectors using a random challenge, or referring to the following rows~\cite{gabizon_proposal_2019}.


\section{Semantic Preservation}
\label{sec:sem-pre-compilation}

Whenever there is a non-obvious transformation from a source representation to a target one, there is a chance that the meaning of the source program which the programmer intended to write will not be preserved perfectly.
In this section we define formally what it means for \clap to preserve the semantics of a circuit when it is converted to a constraint system.
We take advantage of the architecture we presented in the previous sections without which the following definitions would be impossible.
Specifically, we define semantic preservation for \code{gen\_cs} by drawing a connection between high-level semantics \code{gen\_trace} and low-level semantics \code{sat}.


\subsection{Completeness}
We would expect that all the behaviors of the source are present in the target.
In our case, this means that any trace produced by \code{gen\_trace} for a circuit $c$ should be a satisfying assignment for the constraint system obtained from $c$.
This rules out over-constrained systems, where an input that is valid for the circuit is not satisfying for the constraint system.

\begin{definition}{Completeness}\\
  We say that \code{gen\_cs} is complete iff:
  \begin{lstlisting}[mathescape]
  $\forall$ (c : circuit) (input trace : trace).
  gen_trace c input = Some trace
  $\Longrightarrow$ sat (gen_cs c) trace = true
  \end{lstlisting}\vskip -1.5em
  where \code{++} denotes list concatenation.
\end{definition}
Note that there are some harmless cases where constraints are redundant or unnecessary, for example, a duplicated \code{boolCheck} gate for the same wire. We will ignore these cases and talk about over-constraining only for cases where the constraint systems accept strictly fewer witnesses than its corresponding circuit, as defined above.
The completeness of our compilation from circuit to constraint system is a logical continuation of the completeness provided by any proof system between the statement being proved (here encoded as a constraint system) and the resulting proof.


\subsection{Soundness}
More insidious is the case where a new behavior is introduced in the low-level which was not present at the high-level. In this case, any testing or formal verification done at the high-level is of no help.
To avoid this case, we expect that any trace that can satisfy the constraint system can also be generated by \code{gen\_trace}.
This rules out under-constrained systems, which can be satisfied by inputs that were not originally intended by the developer.

\begin{definition}{Soundness}\\
  We say that \code{gen\_cs} is sound iff:
  \begin{lstlisting}[mathescape]
  $\forall$ (c : circuit) (input trace : trace).
  sat (gen_cs c) (input ++ trace) = true
  $\Longrightarrow$ gen_trace c input $\approx_{c}$ Some (input ++ trace)
  \end{lstlisting}\vskip -2em
  where $\approx_{c}$ is a function that matches two traces only on the positions of their inputs and outputs (i.e., it skips intermediate variables), as defined by the circuit.
\end{definition}

Notice that we do not require \code{gen\_trace} to produce a witness that is simply equal to the one universally quantified but only equal up-to the $\approx_{c}$ relation.
The reason is that there could be several traces satisfying the constraint system which are equal in the positions corresponding to the inputs and outputs of the circuits, but different in other intermediate positions.
This is perfectly fine as long as there is a deterministic correspondence between inputs and outputs.
However, \code{gen\_trace} is a deterministic function and therefore can produce only one of such traces.
Relaxing the property with $\approx_{c}$ allows the single witness produced by \code{gen\_trace} to match an equivalence class of witnesses with the same input/output behavior.
In this way we can capture correctly non-fully-constrained witnesses which are necessary to support oracles, as discussed in Section \ref{sec:oracles}.

The soundness of our compilation from circuit to constraint system is a logical continuation of the (probabilistic) soundness provided by any proof system between the statement being proved (here encoded as a constraint system) and the resulting proof.

\subsection{Agda Model}
\label{sec:agda}
Agda is a dependently typed functional programming language\footnote{\url{https://wiki.portal.chalmers.se/agda/pmwiki.php}} and a proof assistant based on intuitionistic type theory.
In our model of \clap we only address the correspondence between \code{gen\_trace} and \code{gen\_cs} functions.
We do not go further down in our compilation stack like tabulation.

The main takeaway is that our proof is \emph{compositional} with respect to any choice of sound and complete sub-circuits.
This implies that a gate designer can prove each gate sound and complete separately and obtain the same properties for any circuit composed by those gates.
The Agda theorems for sequential composition are shown in Figure~\ref{fig:agda}.

\begin{figure}[ht]
  \begin{subfigure}[b]{0.22\textwidth}
\begin{lstlisting}[mathescape]
soundnessSeq :
  ($c_1$ $c_2$ : Circuit)
  -> Sound $c_1$
  -> Sound $c_2$
  -> Sound (seq $c_1$ $c_2$)
\end{lstlisting}
  \end{subfigure}
\begin{subfigure}[b]{0.25\textwidth}
\begin{lstlisting}[mathescape]
completenessSeq :
  ($c_1$ $c_2$ : Circuit)
  -> Complete $c_1$
  -> Complete $c_2$
  -> Complete (seq $c_1$ $c_2$)
\end{lstlisting}
  \end{subfigure}
  \caption{Agda theorems for soundness and completeness of \code{seq}.}
  \label{fig:agda}
\end{figure}

It is worth mentioning two particular features of our system that presented additional complexity in the proofs: assertions and oracles.

The proof of completeness becomes more involved when we go from a purely functional definition of \code{gen\_trace} into the case where the function can fail (e.g. because of an assertion).
This was necessary to deal with even simple gates such as \code{boolCheck}.

On the other hand, the proof of soundness becomes significantly more complex when dealing with non-fully-constrained traces.
In particular, we first defined a notion of \emph{strong} soundness, where we assume that all gates produce fully-constrained traces.
In this case we can simply require equality between the witness satisfying \code{sat} and the one produced by \code{gen\_trace}.
In the presence of oracles, soundness becomes subtle. In particular, we need to identify for each base gate of a circuit which are the positions in a trace that are fully determined by the input.
Relaxing soundness to allow for non-fully-constrained traces required the introduction of the $\approx_c$ equality introduced previously.
This was necessary to deal with oracles gates such as \code{isZero}, as described in Section~\ref{sec:oracles}.

It is important to note that the Agda model only \emph{mimics} the actual  Rust implementation; it is not directly connected to it through any kind of code extraction technique at the moment. It is worth mentioning that in Agda we also implement a variation of builders as monadic computations, which could be used in the future to derive full functional correctness and not just semantic preservation.

\section{Optimizations}
\label{sec:optimizations}

In this section, we describe what optimization passes \clap applies at the circuit level in order to achieve the same (or better) circuit sizes as the hand-optimized versions.

Decoupling the optimization from the circuit implementation not only makes both auditing and verifying easier, but it also tends to find more potential optimization opportunities than what developers can apply manually.
Applying whole-circuit optimizations \emph{after} the circuit is built has the advantage that the optimizer can carefully introduce locally unsound gates, which enables us to safely remove duplicated checks and inline across function boundaries.

It should be noted that, while some optimizations could be carried out on the constraint system, applying them at the circuit level has the advantage that the witness generator is also optimized.

\subsection{Arithmetic Inlining}
\label{sec:inlining}
The main role of the circuit data structure is to introduce and manage intermediate variables that can be easily manipulated in an eDSL.
Additionally, writing simple and readable circuits can lead to sparse gates i.e. gates where wires are unused or constants are zero.
However, once the circuit is completed, many of those intermediate variables can be removed by \emph{inlining} them, leading to gates used to the fullest.

\begin{figure}
\centering
\[
\inferrule[LinearInliningOnL]
{\w{x_1} \coloneqq c_0\cdot\w{x_0} + c_1 \\\\
 \w{x_3} \coloneqq ql\cdot\w{x_1} + qr\cdot\w{x_2} + qm\cdot(\w{x_1}\cdot\w{x_2)} + qc }
{\w{x_3}\coloneqq (ql \cdot c_1) \cdot \w{x_0} + (qr + qm \cdot c_2)\cdot \w{x_2} \\\\
\hspace*{2em} + (qm \cdot c_1) \cdot \w{x_0}\cdot \w{x_2} + (ql \cdot c_2 + qc)}
\]

\[
\inferrule[BooleanReduction]
{ \w{x} \coloneqq ql \cdot \w{b} + qm \cdot \w{b}^2 + qc  \\ \w{b} \in \{0, 1\}}
{\w{x} \coloneqq (ql + qm)\cdot \w{b} + qc}
\]

\[
\inferrule[ConversionToLC]
{\w{x_1} \coloneqq q_0 \cdot \w{y_0} \\
 \w{x_2} \coloneqq q_1 \cdot \w{y_1} + \w{x_1}\\\\
 \w{x_3} \coloneqq q_2 \cdot \w{y_2} + \w{x_2}}
{\w{x_3} \coloneqq \Sigma_{i=0}^2 q_i\cdot \w{y_i}}
\]
    \caption{Examples of optimizations applied by \clap.}
    \label{fig:optimizations}
\end{figure}

The main limitation of inlining is respecting the degree and shape of the gates.
For example, for the classic PlonK \code{arith} gate we only have right to degree two and we can use a single multiplication.

\point{Linear inlining}
However, as shown in~\cite{ambrona_new_2022} and confirmed by our experiments in Section~\ref{sec:experiments}, \emph{linear} inlining is already very effective for circuits like Poseidon. The optimization pass \textsc{LinearInliningOnL} from Figure~\ref{fig:optimizations} describes one of such passes, where a linear term can be inlined into the left wire of an arithmetic gate.

\point{Boolean inlining}
Additionally, the knowledge that a variable is boolean can unlock more inlining opportunities because its degree does not increase (i.e. $0^2=0$, $1^2=1$).
A simple example of such an optimization is the \textsc{BooleanReduction} rule, which removes the squaring of boolean wires.

\point{Circuit flattening}
\cite{gabizon_plonk_2019} introduced the notion of custom gate which, in \clap terms, is a new kind of constrained vector that requires new identities.
Custom gates allow to dramatically reduce the size of a circuit by encoding a potentially large computation as a single gate, provided that the degree is not too large.
A smaller circuit requires less proving time but each custom gate adds to verification time and proof size, a trade-off that is sensible if the gate is used extensively.

To build a custom gate from a circuit description, we need to inline the individual gates it is composed of in a set of polynomial equations that will form the definition of our new custom gate.
This process can be automated by passing the original circuit description to the inliner and instructing it to inline as much as possible while only respecting the degree.
If the inliner is not limited by a particular gate shape, it produces polynomial equations that are as long as possible, and it breaks them only to avoid exceeding the degree.

Using flattening, we can reuse existing audited circuit definitions to automatically derive custom gates. Now that the process does not require long manual re-implementations, it allows more experimentations of what parts of a circuit are worth transforming to custom gates.
In our experiments in Section~\ref{sec:experiments}, we demonstrate flattening on a very large gate performing a Poseidon round and achieve the same results as the manual implementation.

\subsection{Gate Conversion}
Another useful transformation consists in converting one or more gates to a different one that could be more compact or simply the only one available in a target system.
For example \clap builder always produces circuits that use the \code{arith}, making them suitable for any classic PlonK system.
Boojum on the other hand, does not support the \code{arith} gate and prefers more specialized gates to perform Fused-Addition-Multiplication and Linear-Combinations.
\clap gate conversion can easily transform \code{arith} into \code{FMA} when producing Boojum circuits or can leave them unchanged when targeting classic PlonK. This transformation is captured by the \textsc{ConversionToLC} optimization pass defined in Figure~\ref{fig:optimizations}.

More interestingly, when \clap detects several \code{arith} gates chained in a linear way, it replaces them with a linear combination gate, saving many intermediate variables.

\subsection{Taming Locally-unsound Gates}

\point{Common sub-expression elimination}
There are many occasions where the same expression is used multiple times in a circuit, and it greatly simplifies the builder to simply copy it rather than keeping a cache of already computed expressions.
During the common sub-expression elimination pass, the optimizer finds variables that are the output of the same gate and keeps only one, significantly reducing duplication and unnecessary intermediate variables.

\point{Duplicate range-checks}
A special case of common sub-expression elimination is when there are multiple range-checks over the same variable.
For example, a gate that receives as input or returns as output a boolean variable must range-check it in order to be individually safe to use.
However, if we chain two such gates, we end up with two boolean checks over the same variable.
This problem occurs very often when working with integers of 32 or 64 bits, as they are usually represented by limbs which need to be individually range-checked.
In our experiments, for example, removing these duplicate checks is crucial to achieve a practical implementation of SHA2-256, which brings us back to the \code{split\_36\_bits\_unchecked} function.

A \clap developer has only access to a sound and complete gate \code{split\_36\_bits} which does introduce all the necessary, and possibly redundant, range-checks.
However, the optimizer has access to the gate \code{split\_36\_bits\_unchecked}, which is unsound in isolation but is perfectly sound when accompanied by another gate that constrains its variables.

The compositional result presented in Section~\ref{sec:agda}, which states that circuits composed of sound and complete sub-circuits are, in turn, sound and complete, cannot be applied directly to unsound gates.
However, as long as the optimizer makes sure unsound gates are used safely, we can still prove soundness for a sub-circuit.
Once we have a sound sub-circuit, our compositional result applies.

The crucial difference in our approach is that unsafe gates can only be introduced by the optimizer, after the builder has produced a safe circuit.
In contrast, when an unsound gate is used in Boojum, the responsibility of its safe use falls on the developer's shoulders.

\subsection{Semantic Preservation}

We now look at how we can ensure that each optimization preserves the semantics of the original circuit written by the developer.
Considering that the semantics of a circuit is defined as its trace generator, we need to make sure that witnesses produced before and after each optimization match.

In order to ensure preservation, each optimization should respect the following property.
\begin{definition}{Semantic preservation}\\
  We say that an optimization \code{opt} is semantic-preserving iff:
  \begin{lstlisting}[mathescape]
  $\forall$ (c : circuit) (input : trace).
  gen_trace (opt c) input $\approx_c$ gen_trace c input
  \end{lstlisting}
\end{definition}\vskip -1.5em

Given that optimizations can (and should!) remove intermediate variables, we don't require the resulting witnesses to be equal, but simply to correspond in their input-output positions (denoted $\approx_c$ as for soundness in Section~\ref{sec:sem-pre-compilation}).
Notice that, just as importantly as the case when a witness is returned, we need to ensure that trace generation fails for the same inputs before and after an optimization, i.e. that in both cases \code{gen\_trace} returns \code{None}.
This means, for example, that an optimization has not accidentally removed an assertion.

%
%
%
Proving this property for the optimizations currently implemented in \clap is left for future work.

\section{Experimental Evaluation}
\label{sec:experiments}
As explained in Section~\ref{sec:introduction}, one of \clap’s core tenets is to enable the writing of circuits, relying on an automatic optimizer to make them efficient. This section presents the experiments that support this approach, assessing both the qualitative simplicity and safety of the circuit definitions and the quantitative efficiency of the automatically optimized ones.

\subsection{A Note on Boojum}

The experimental evaluation uses Boojum circuits as a point of comparison.
Boojum circuits are used in production by ZKsync Era, are developed and manually optimized by experts, and are reviewed by third party auditors.
Circuits in Boojum are written in a Rust eDSL by which \clap was partly inspired. Boojum is a FRI-based STARK proving system that uses Plonkish arithmetization. Boojum supports recursion, which is heavily used for ZKsync proofs. The recursion-centric approach means that (a) in-circuit hashing (Poseidon2~\cite{grassi_poseidon2_2023}) becomes crucial for the system and (b) verification time and proof size have lesser significance, which in turn means that the low-level table can be wide. Due to these two observations, Boojum introduces a custom gate that expresses an entire round of Poseidon2.

\begin{figure}[tb]
\begin{subfigure}{0.45\textwidth}
\begin{lstlisting}[language=Rust]
fn xor(&mut self,
       l: Repr<F, bool>,
       r: Repr<F, bool>) -> Repr<F, bool> {
  // (l /\ ~r) \/ (~l /\ r)
  let nr = self.not(r);
  let lnr = self.and(l, nr);
  let nl = self.not(l);
  let nlr = self.and(nl, r);
  self.or(lnr, nlr)
}
\end{lstlisting}
\caption{\clap version.}
\label{fig:clap-xor}
\end{subfigure}
\begin{subfigure}{0.45\textwidth}
\begin{lstlisting}[language=Rust]
pub fn xor(self, cs: &mut CS, other: Self) -> Self {
    // c = ~(a /\ b) /\ ~(~a /\ ~b)
    // =(1 - (a * b)) * (1 - ((1 - a) * (1 - b)))
    // =(1 - ab) * (1 - (1 - a - b + ab))
    // =(1 - ab) * (a + b - ab)
    // =a + b - ab - (a^2)b - (b^2)a + (a^2)(b^2)
    // =a + b - ab - ab - ab + ab
    // =a + b - 2ab
    // -2a * b = c - a - b
    // 2a * b = a + b - c
    // 2a*b - a - b + c = 0
    let a_plus_b = compute_fma(cs, ...);
    let c = compute_fma(cs,...);
    Boolean::from_variable_checked(cs, c)
}
\end{lstlisting}
\caption{Boojum version.}
\label{fig:boojum-xor}
\end{subfigure}
\caption{Implementation of XOR.}
\label{fig:xor}
\end{figure}

\subsection{Safe and Simple Circuits}

The initial comparison point is the implementation of the \code{xor} function in both systems.
The \clap implementation, shown in Figure~\ref{fig:clap-xor}, follows its textbook definition. In contrast, the Boojum version in Figure~\ref{fig:boojum-xor} is manually explained with a sequence of comments starting from a simple identity until it reaches an equivalent form that is compatible with a supported gate. This implementation is error-prone and hard to audit, as a reviewer would have to check both the \code{xor} logic and the manual optimization at the same time. The \clap optimizer is able to produce the exact same circuit from the simpler definition, thanks to the Boolean inlining optimizations presented in Section~\ref{sec:optimizations}.

\begin{figure}[tb]
\begin{subfigure}{0.45\textwidth}
\begin{lstlisting}[language=Rust]
pub fn round_function<F: SmallField>(
    cs: &mut Env<F>, state: &mut State<F>,
    message_block: &[Repr<F, u32>; 16],
    range_check_final_state: bool,
) -> Option<[Repr<F, U4>; 64]> {
\end{lstlisting}
\caption{\clap version.}
\label{fig:clap-sha-signature}
\end{subfigure}
\begin{subfigure}{0.45\textwidth}
\begin{lstlisting}[language=Rust]
// we only have variables in the signature:
// - if it's a first round then state is made of constants
// - if it's not the last round then next round will range check when rounds happen
// - if it's the last round then it's done by caller
pub fn round_function<F: SmallField>(
    cs: &mut CS, state: &mut [Variable; 8],
    message_block: &[Variable; 16],
    range_check_final_state: bool,
) -> Option<[Variable; 64]> {
\end{lstlisting}
\caption{Boojum version.}
\label{fig:boojum-sha-signature}
\end{subfigure}
\caption{Signatures for SHA round function.}
\label{fig:sha-signature}
\end{figure}

The following examples analyze portions of the implementation of the SHA-256 hash function, starting with the function signature for a SHA round. Figure~\ref{fig:boojum-sha-signature} presents both Boojum and \clap  versions of it. Boojum's \code{round\_function} is not precisely typed to avoid redundant range checks, as justified in the comment. In contrast, the \clap version uses precise type—safe range validation, while its optimizer eliminates duplicate checks, thus improving both clarity and safety.

\begin{figure}[tb]
\begin{subfigure}{0.45\textwidth}
\begin{lstlisting}[language=Rust]
for idx in 16..SHA256_ROUNDS {
  ...
  let (u32_part, _) = range_check_36(cs, expanded_word);
  expanded[idx] = u32_part;
}
\end{lstlisting}
\caption{\clap version.}
\label{fig:clap-sha-unsound}
\end{subfigure}
\begin{subfigure}{0.45\textwidth}
\begin{lstlisting}[language=Rust]
// we assume inputs to be range checked, so
// idx - 15 is range checked by xor
// idx - 2 is range-checked by xor
// so we are "safe" as long as our index over which we work right now will become
// idx - 2 for some next round
for idx in 16..SHA256_ROUNDS {
  ...
  // we only need to fully range check if it's one of the last
  let u32_part = if idx + 2 >= 64 {
    let (u32_part, _) = range_check_36(cs, expanded_word);
    u32_part
    } else {
      let (u32_part, high_unchecked) = split_36_bits_unchecked(cs, expanded_word);
      yet_unconstrained_chunks.push(high_unchecked);
      u32_part
  };
  expanded[idx] = u32_part;
}
// range check small pieces
...
\end{lstlisting}
\caption{Boojum version.}
\label{fig:boojum-sha-unsound}
\end{subfigure}
\caption{Unsound functions in SHA.}
\label{fig:sha-unsound}
\end{figure}

\begin{figure}[tb]
\centering
\setlength{\tabcolsep}{4pt}
\begin{tabular}{|l|rrr|}
\hline
    \bf Optimization kind &\bf XOR  &\bf Poseidon2  &\bf SHA-256 \bigstrut \\
    \hline\hline
    Arithmetic inlining &- &1,324  &64 \bigstrut[t] \\
    Boolean inlining &3 &-  &- \\
    Common-subexpression el. &- &237  &286 \\ 
    Common-range-check el. &- &-  &4,298 \\ 
    Gate replacement &- &675  &64 \\
    \hline
\end{tabular}
\caption{Frequency of optimization-pass application.}
\label{tab:optimization-freq}
\end{figure}

\begin{figure*}[tb]
  \centering
  \begin{subfigure}{\textwidth}
    \centering
    \setlength{\tabcolsep}{10pt}
    \begin{tabular}{|l|rrr|r|}
      \hline
      &\bf Constant  &\bf FMA  &\bf Linear Combination (size 4) &\bf Total \bigstrut \\
      \hline\hline
      Boojum non-custom-gate  &5 &854 &304 &1,163 \bigstrut\\
      \clap non-optimized &239  &1,874  &0 &2,113\\
      \clap optimized &1  &602  &451 &1,054\\
      \hline
    \end{tabular}
    \caption{Circuit sizes (per constraint) for Poseidon2 circuit implementations.}
    \label{tab:poseidon2}
  \end{subfigure}\\
  \begin{subfigure}{\textwidth}
    \centering
    \setlength{\tabcolsep}{7pt}
    \begin{tabular}{|l|rrrr|r|}
        \hline
         &\bf Constant  &\bf FMA  &\bf Linear Combination (size 4) &\bf Lookup &\bf Total \bigstrut \\
        \hline\hline
        Boojum   &76 &1,268 &3,617 &3,885 &8,845 \bigstrut \\
        \clap non-optimized &84  &2,434  &4,832 &6,802 &14,152\\
        \clap optimized &77  &995  &3,245 &3,596 &7,913\\
        \hline
    \end{tabular}
    \caption{Circuit sizes (per constraint) for (one round of) SHA-256 circuit implementations.}
    \label{tab:sha2}
  \end{subfigure}
    
  \caption{Optimization effects measured.}
\end{figure*}

    \begin{figure}[tb]
    \centering
    \setlength{\tabcolsep}{3pt}
    \begin{tabular}{|l|rrrr|}
    \hline
        \bf Poseidon2 version &\bf Rows  &\bf Width  &\bf Witness size &\bf Degree \bigstrut \\
        \hline\hline
        \clap optimized &1,054 &5 &1,066 &5 \bigstrut\\
        \clap flattened &1  &130 &130 &8\\
        \hline
    \end{tabular}
    \caption{Effect of flattening the Poseidon2 circuit.}
    \label{tab:flatten}
  \end{figure}

The evaluation continues with the ``expansion'' phase of SHA-256, which operates over \code{u32}. Figure~\ref{fig:sha-unsound} shows code snippets from both \clap  and Boojum. This phase generates an expanded word expected to fit within~36 bits, with the intention to keep and verify just the lower~32 bits.

Boojum's implementation, as previously introduced in Section~\ref{sec:motivation}, uses the unsafe function \code{split\_36\_bits\_unchecked} for most rounds to handle the expanded words, while collecting high \code{u4}s for batched range checks later. The final two rounds are immediately range-checked as their outputs are not utilized in subsequent rounds. In contrast, \clap has to apply comprehensive~36-bit range-checks to all expanded words, with its optimizer removing any redundant checks. This approach results in a significantly simpler implementation and reduces the developer's mental burden of tracking these ranges manually.

It is also important to highlight the experimentation with automatic derivation of custom gates. Boojum's implementation of the Poseidon2 custom gate spans over~800 lines of code, featuring the Poseidon2 logic twice (once for evaluation and once for witness generation). In contrast, \clap mechanically derives its version from the naive implementation through the circuit flattening optimization described in Section~\ref{sec:optimizations}.

\subsection{Hand-Optimized Performance, and Beyond}

In this section, we compare the performance for both Boojum and \clap circuits, in terms of the number of constraints, as it is a metric proportional to proving times. However, the exact mapping of circuit size to proving time depends on various factors such as table geometry, levels of parallelism, etc. Unless explicitly mentioned, other parameters such as gate widths are equal in both systems. Our experimentation was mostly focused on the Poseidon2 and SHA-256 hash functions. These circuits are compiled targeting a subset of Boojum's constrained vectors: Fused-Multiply-Add~(FMA) without constant (\code{c0 * a * b + c1 * c - d = 0}), constant allocation, linear combination, and lookups.
Our focus is on evaluating \clap's optimizations; Table~\ref{tab:optimization-freq} presents a summary of how often each of the optimization passes described in Section~\ref{sec:optimizations} are applied.

\point{Poseidon2, non-custom gate version}
There are two versions of Poseidon2 in Boojum. The first is a naive implementation without custom gates. The \clap translation of this circuit is almost identical to the Boojum one, with the only difference being that it expresses all operations using the \code{arith} gate, thus building a circuit which could be compiled to classic PlonK in a straightforward way.
However, instead of compiling directly, the \code{arith} gates are treated as an intermediate representation that can be compiled to the Boojum constrained vectors. As shown in Figure~\ref{tab:optimization-freq}, the main source of optimization for this circuit is the inlining of arithmetic expressions and replacing them with new linear combination gates.
Figure~\ref{tab:poseidon2} shows the circuit size comparison for Poseidon2 between Boojum and \clap.
The unoptimized \clap implementation is~80\% larger than the one from Boojum.
However, after the optimization pass, it ends up being almost~10\% smaller.

\point{Poseidon2, flattened version}
The second implementation of Poseidon2 is the custom gate mentioned above produced by flattening~\ref{sec:inlining}.
Figure~\ref{tab:flatten} shows the effect of flattening an entire round of Poseidon2 into one custom gate: the key observation is that it reduces the number of constraints, or rows in the low-level table, to just~1, which in turn decreases the proving time. Witness size is also reduced, as there are only~130 variables in the flattened version, which makes witness generation faster. On the flip side, the width of the low-level table and the degree of the polynomials are increased. The first increases the verification time and proof size, while the second increases the proof time. Automating the process of flattening circuits can allow the circuit developer to quickly experiment to find the right trade-off between these dimensions.

\point{SHA-256}
The evaluation concludes with specific metrics for the SHA-256 circuit. The naive circuit exhibits a~60\% increase in size compared to the Boojum hand-optimized version, as shown in Figure~\ref{tab:sha2}. Upon application of the optimizer, \clap produces a circuit that is~10\% smaller and offers increased safety guarantees. The main optimization pass in this case
was the elimination of duplicated range-checks, expected from \clap's restriction to only allow sound gates to be used by the circuit developer.

\section{Related Work}
\label{sec:related}

The closest work to \clap are Rust eDSLs like Boojum and Halo2.
Indeed, \clap tries to be a principled breakdown of the same approach.
The main difference is that in these projects it is difficult to identify the different semantics layers and gates are developed ``all at once'' going from witness generator directly to table layout.
Moreover, the proof system is part of the same code base, and there is no point in the code structure where the constraint system can be extracted, for example, to use it with a different proof system.
Lastly, these projects offer no guarantee of semantic preservation on their own, there are, however, separate efforts.
\cite{soureshjani_automated_2023} proposes static analysis techniques over Halo2 constraint systems, backed by SMT.
While these analyses can help discover soundness bugs, they are incomplete, and the experimental evaluation is limited.

\cite{certik_verification_2024} formally verifies in Coq a zkWASM implementation in Halo2.
The formalization covers soundness of the constraint system with respect to a specification of WASM language.
While an impressive achievement, this work is an example of ex-post-facto verification, that does not integrate in the workflow of developers.
Compared to \clap, it provides additionally functional correctness with respect to the WASM specification, but it lacks completeness for the witness generators.

o1-js~\cite{mina_o1-js_2024} offers a circuit eDSL in JavaScript while Noir~\cite{aztec_noir_2024} is a full language stack.
Both projects do not allow the developer to define custom gates or to influence the tabulation strategy beyond what is supported by their specific proof systems.
They also do not offer any formal correctness guarantees.

The following systems do not produce Plonkish constraint systems but provide relevant context.\\
\indent Circom \cite{belles-munoz_circom_2022} is a hardware description language for R1CS and, being one of the first, has a large use base.
However, it has the downsides of standalone languages and offers a very low-level programming experience to the user.
The Circomspect project applies static analysis to Circom circuits~\cite{trail_of_bits_circomspect_2024}.
Coda offer a very similar approach in a more principled way using the Coq proof assistant~\cite{liu_certifying_2023}.\\
\indent Leo~\cite{chin_leo_2021} is a high-level general-purpose language for zero-knowledge applications on the Aleo blockchain. It exclusively targets R1CS arithmetization and completeness is proven in ACL2.\\
\indent Cairo~\cite{goldberg_cairo_2021}, similarly to Leo, is another high-level language for ZK circuits. It targets a low-level VM, the Cairo CPU, for which an AIR arithmetization is provided. A formalization of the Cairo CPU and it's AIR representation has been done in Lean~\cite{avigad_verified_2022}.\\
\indent ZoKrates~\cite{eberhardt_zokrates_2018} introduced a model of off-chain computation based on zk-SNARKS and included one of the first modern DSL targeting R1CS, although without any formal guarantee.\\
\indent CirC~\cite{ozdemir_circ_2022} is a compiler infrastructure for multiple front-ends and back-ends, including R1CS, with makes heavy use of SMTs. It supports sophisticated optimizations and some support for bug finding, but no formal guarantees of correctness.


\section{Future Work}
\label{sec:future}

A natural continuation of this work would be completing the mechanization of our formalization in a proof-assistant.
In particular we would like to prove soundness and completeness for a representative set of Boojum gates and semantic preservation for some of the most effective optimizations.

In addition to manually encoding circuit optimizations, we can also envision automatic optimization synthesis, as has been proposed for compilers~\cite{lopes_weakest_2014}.

In addition to improving \clap, we also envision building on top of it. In particular, once we have semantic preservation from the circuit down to the proof, we can start to work on the functional correctness of the circuits themselves.
Concretely, we plan on writing a new builder library in a proof-assistant and reuse the rest of our pipeline.
In the proof assistant, we could prove the equivalence between a circuit and its functional specification.

A crucial circuit of the ZKsync verification stack is the recursive FRI verifier that aggregates a large number of EVM proofs in a single one. Verifying the functional specification of this circuit using \clap embedded in a proof assistance would be a first logical step and it would fit into the larger plan to formally verify the SNARK verification chain of ZKsync Era. Verifying the verifier opens up the possibility of decentralized proof production~\cite{firsov_ouroboros_2024}.

Lastly, it would be interesting to explore whether our approach extends to other constraints systems such as Air or CCS.

\section{Conclusions}
\label{sec:conclusions}

\clap proposes the first circuit compiler architecture for Plonkish capable of producing constraint systems and witness generators that are sound and complete by construction.
Having multiple steps is crucial for defining and proving properties of semantic preservation, which are increasingly needed as proof systems handle statements of increasing complexity.
Optimizations, when applied at the right level of abstraction, allow for simpler, safer and ultimately more reusable circuit definitions, while providing better results than hand-written circuits.

\bibliographystyle{alpha}
\bibliography{main}

\newcommand{\etalchar}[1]{$^{#1}$}
\begin{thebibliography}{BMIMT{\etalchar{+}}23}

\bibitem[AGL{\etalchar{+}}22]{avigad_verified_2022}
Jeremy Avigad, Lior Goldberg, David Levit, Yoav Seginer, and Alon Titelman.
\newblock A verified algebraic representation of cairo program execution.
\newblock In {\em Proceedings of the 11th {ACM} {SIGPLAN} {International}
  {Conference} on {Certified} {Programs} and {Proofs}}, pages 153--165,
  Philadelphia PA USA, January 2022. ACM.

\bibitem[ASTW22]{ambrona_new_2022}
Miguel Ambrona, Anne-Laure Schmitt, Raphael~R. Toledo, and Danny Willems.
\newblock New optimization techniques for {PlonK}’s arithmetization, 2022.
\newblock Publication info: Preprint. MINOR revision.

\bibitem[{Azt}24]{aztec_noir_2024}
{Aztec}.
\newblock Noir {Language}, April 2024.

\bibitem[BGH19]{bowe_recursive_2019}
Sean Bowe, Jack Grigg, and Daira Hopwood.
\newblock Recursive {Proof} {Composition} without a {Trusted} {Setup}, 2019.
\newblock Publication info: Preprint. MINOR revision.

\bibitem[BMIMT{\etalchar{+}}22]{belles-munoz_circom_2022}
Marta Belles-Munoz, Miguel Isabel, Jose~Luis Munoz-Tapia, Albert Rubio, and
  Jordi Baylina.
\newblock Circom: {A} {Circuit} {Description} {Language} for {Building}
  {Zero}-knowledge {Applications}.
\newblock {\em IEEE Trans. Dependable and Secure Comput.}, pages 1--18, 2022.

\bibitem[BMIMT{\etalchar{+}}23]{belles-munoz_circom_2023}
Marta Bellés-Muñoz, Miguel Isabel, Jose~Luis Muñoz-Tapia, Albert Rubio, and
  Jordi Baylina.
\newblock Circom: {A} {Circuit} {Description} {Language} for {Building}
  {Zero}-{Knowledge} {Applications}.
\newblock {\em IEEE Transactions on Dependable and Secure Computing},
  20(6):4733--4751, November 2023.
\newblock Conference Name: IEEE Transactions on Dependable and Secure
  Computing.

\bibitem[BSBHR18]{ben-sasson_scalable_2018}
Eli Ben-Sasson, Iddo Bentov, Yinon Horesh, and Michael Riabzev.
\newblock Scalable, transparent, and post-quantum secure computational
  integrity, 2018.
\newblock Publication info: Preprint. MINOR revision.

\bibitem[CBBZ23]{chen_hyperplonk_2023}
Binyi Chen, Benedikt Bünz, Dan Boneh, and Zhenfei Zhang.
\newblock {HyperPlonk}: {Plonk} with {Linear}-{Time} {Prover} and
  {High}-{Degree} {Custom} {Gates}.
\newblock In Carmit Hazay and Martijn Stam, editors, {\em Advances in
  {Cryptology} - {EUROCRYPT} 2023 - 42nd {Annual} {International} {Conference}
  on the {Theory} and {Applications} of {Cryptographic} {Techniques}, {Lyon},
  {France}, {April} 23-27, 2023, {Proceedings}, {Part} {II}}, volume 14005 of
  {\em Lecture {Notes} in {Computer} {Science}}, pages 499--530. Springer,
  2023.

\bibitem[{Cer}24]{certik_verification_2024}
{CertiK}.
\newblock Verification of {zkWasm} in {Coq}, June 2024.

\bibitem[CET{\etalchar{+}}24]{chaliasos_sok_2024}
Stefanos Chaliasos, Jens Ernstberger, David Theodore, David Wong, Mohammad
  Jahanara, and Benjamin Livshits.
\newblock {SoK}: {What} don't we know? {Understanding} {Security}
  {Vulnerabilities} in {SNARKs}.
\newblock {\em CoRR}, abs/2402.15293, 2024.
\newblock arXiv: 2402.15293.

\bibitem[CWC{\etalchar{+}}21]{chin_leo_2021}
Collin Chin, Howard Wu, Raymond Chu, Alessandro Coglio, Eric McCarthy, and Eric
  Smith.
\newblock Leo: {A} {Programming} {Language} for {Formally} {Verified},
  {Zero}-{Knowledge} {Applications}.
\newblock {\em IACR Cryptol. ePrint Arch.}, page 651, 2021.

\bibitem[ET18]{eberhardt_zokrates_2018}
Jacob Eberhardt and Stefan Tai.
\newblock {ZoKrates} - {Scalable} {Privacy}-{Preserving} {Off}-{Chain}
  {Computations}.
\newblock In {\em {IEEE} {International} {Conference} on {Internet} of {Things}
  ({iThings}) and {IEEE} {Green} {Computing} and {Communications} ({GreenCom})
  and {IEEE} {Cyber}, {Physical} and {Social} {Computing} ({CPSCom}) and {IEEE}
  {Smart} {Data} ({SmartData}), {iThings}/{GreenCom}/{CPSCom}/{SmartData} 2018,
  {Halifax}, {NS}, {Canada}, {July} 30 - {August} 3, 2018}, pages 1084--1091.
  IEEE, 2018.

\bibitem[FL24]{firsov_ouroboros_2024}
Denis Firsov and Benjamin Livshits.
\newblock The {Ouroboros} of {ZK}: {Why} {Verifying} the {Verifier} {Unlocks}
  {Longer}-{Term} {ZK} {Innovation}, 2024.
\newblock Publication info: Preprint.

\bibitem[GKS23]{grassi_poseidon2_2023}
Lorenzo Grassi, Dmitry Khovratovich, and Markus Schofnegger.
\newblock Poseidon2: {A} {Faster} {Version} of the {Poseidon} {Hash}
  {Function}.
\newblock In Nadia El~Mrabet, Luca De~Feo, and Sylvain Duquesne, editors, {\em
  Progress in {Cryptology} - {AFRICACRYPT} 2023}, pages 177--203, Cham, 2023.
  Springer Nature Switzerland.

\bibitem[GPR21]{goldberg_cairo_2021}
Lior Goldberg, Shahar Papini, and Michael Riabzev.
\newblock Cairo - a {Turing}-complete {STARK}-friendly {CPU} architecture.
\newblock {\em IACR Cryptol. ePrint Arch.}, page 1063, 2021.

\bibitem[GW19]{gabizon_proposal_2019}
Ariel Gabizon and Zachary~J Williamson.
\newblock Proposal: {The} {Turbo}-{PLONK} program syntax for specifying {SNARK}
  programs, 2019.

\bibitem[GW20]{gabizon_plookup_2020}
Ariel Gabizon and Zachary~J. Williamson.
\newblock plookup: {A} simplified polynomial protocol for lookup tables, 2020.
\newblock Publication info: Preprint. MINOR revision.

\bibitem[GWC19]{gabizon_plonk_2019}
Ariel Gabizon, Zachary~J. Williamson, and Oana Ciobotaru.
\newblock {PLONK}: {Permutations} over {Lagrange}-bases for {Oecumenical}
  {Noninteractive} arguments of {Knowledge}, 2019.
\newblock Publication info: Preprint.

\bibitem[KPV22]{kattis_redshift_2022}
Assimakis~A. Kattis, Konstantin Panarin, and Alexander Vlasov.
\newblock {RedShift}: {Transparent} {SNARKs} from {List} {Polynomial}
  {Commitments}.
\newblock In {\em Proceedings of the 2022 {ACM} {SIGSAC} {Conference} on
  {Computer} and {Communications} {Security}}, pages 1725--1737, Los Angeles CA
  USA, November 2022. ACM.

\bibitem[LKL{\etalchar{+}}23]{liu_certifying_2023}
Junrui Liu, Ian Kretz, Hanzhi Liu, Bryan Tan, Jonathan Wang, Yi~Sun, Luke
  Pearson, Anders Miltner, Isil Dillig, and Yu~Feng.
\newblock Certifying {Zero}-{Knowledge} {Circuits} with {Refinement} {Types}.
\newblock {\em IACR Cryptol. ePrint Arch.}, page 547, 2023.

\bibitem[LM14]{lopes_weakest_2014}
Nuno~P. Lopes and José Monteiro.
\newblock Weakest {Precondition} {Synthesis} for {Compiler} {Optimizations}.
\newblock In Kenneth~L. McMillan and Xavier Rival, editors, {\em Verification,
  {Model} {Checking}, and {Abstract} {Interpretation} - 15th {International}
  {Conference}, {VMCAI} 2014, {San} {Diego}, {CA}, {USA}, {January} 19-21,
  2014, {Proceedings}}, volume 8318 of {\em Lecture {Notes} in {Computer}
  {Science}}, pages 203--221. Springer, 2014.

\bibitem[{Mat}23]{matter_labs_boojum_2023}
{Matter Labs}.
\newblock Boojum proving system, 2023.

\bibitem[{Min}24]{mina_o1-js_2024}
{Mina}.
\newblock o1-js, April 2024.

\bibitem[OBW22]{ozdemir_circ_2022}
Alex Ozdemir, Fraser Brown, and Riad~S. Wahby.
\newblock {CirC}: {Compiler} infrastructure for proof systems, software
  verification, and more.
\newblock In {\em 2022 {IEEE} {Symposium} on {Security} and {Privacy} ({SP})},
  pages 2248--2266, San Francisco, CA, USA, May 2022. IEEE.

\bibitem[OWBB23]{ozdemir_bounded_2023}
Alex Ozdemir, Riad~S. Wahby, Fraser Brown, and Clark~W. Barrett.
\newblock Bounded {Verification} for {Finite}-{Field}-{Blasting} - {In} a
  {Compiler} for {Zero} {Knowledge} {Proofs}.
\newblock In Constantin Enea and Akash Lal, editors, {\em Computer {Aided}
  {Verification} - 35th {International} {Conference}, {CAV} 2023, {Paris},
  {France}, {July} 17-22, 2023, {Proceedings}, {Part} {III}}, volume 13966 of
  {\em Lecture {Notes} in {Computer} {Science}}, pages 154--175. Springer,
  2023.

\bibitem[PCW{\etalchar{+}}23]{pailoor_automated_2023}
Shankara Pailoor, Yanju Chen, Franklyn Wang, Clara Rodríguez-Núñez,
  Jacob~Van Geffen, Jason Morton, Michael Chu, Brian Gu, Yu~Feng, and Isil
  Dillig.
\newblock Automated {Detection} of {Under}-{Constrained} {Circuits} in
  {Zero}-{Knowledge} {Proofs}.
\newblock {\em Proc. ACM Program. Lang.}, 7(PLDI):1510--1532, 2023.

\bibitem[Set20]{setty_spartan_2020}
Srinath Setty.
\newblock Spartan: {Efficient} and {General}-{Purpose} {zkSNARKs} {Without}
  {Trusted} {Setup}.
\newblock In Daniele Micciancio and Thomas Ristenpart, editors, {\em Advances
  in {Cryptology} – {CRYPTO} 2020}, pages 704--737, Cham, 2020. Springer
  International Publishing.

\bibitem[SHAJ{\etalchar{+}}23]{soureshjani_automated_2023}
Fatemeh~Heidari Soureshjani, Mathias Hall-Andersen, MohammadMahdi Jahanara,
  Jeffrey Kam, Jan Gorzny, and Mohsen Ahmadvand.
\newblock Automated {Analysis} of {Halo2} {Circuits}, 2023.
\newblock Publication info: Published elsewhere. Satisfiability Modulo Theories
  2023 (SMT 2023).

\bibitem[{Tra}24]{trail_of_bits_circomspect_2024}
{Trail of Bits}.
\newblock Circomspect, April 2024.
\newblock original-date: 2022-05-27T13:26:28Z.

\bibitem[{ZCa}22]{zcash_halo2_2022}
{ZCash}.
\newblock Halo2, 2022.

\end{thebibliography}

\clearpage
\begin{appendices}
\section{Further comparison of Clap and Boojum circuit definitions}
\label{sec:appendix}

\begin{figure}[ht]
  \centering
  \begin{subfigure}{0.35\textwidth}
    \centering
    \begin{lstlisting}[language=Rust]
state[0] = reduce_terms(cs,
            [F::TWO, F::ONE, F::ONE],
            [x0, x4, x8]);
state[1] = reduce_terms(cs,
            [F::TWO, F::ONE, F::ONE],
            [x1, x5, x9]);
state[2] = reduce_terms(cs,
            [F::TWO, F::ONE, F::ONE],
            [x2, x6, x10]);
state[3] = reduce_terms(cs,
            [F::TWO, F::ONE, F::ONE],
            [x3, x7, x11]);
    \end{lstlisting}
    \caption{\clap version.}
    \label{fig:clap-hclc-full}
  \end{subfigure}
  \begin{subfigure}{0.35\textwidth}
    \centering
    \begin{lstlisting}[language=Rust]
state[0] = ReductionGate::reduce_terms(
    cs,
    [F::TWO, F::ONE, F::ONE, F::ZERO],
    [x0, x4, x8, zero]
);
state[1] = ReductionGate::reduce_terms(
    cs,
    [F::TWO, F::ONE, F::ONE, F::ZERO],
    [x1, x5, x9, zero]
);
state[2] = ReductionGate::reduce_terms(
    cs,
    [F::TWO, F::ONE, F::ONE, F::ZERO],
    [x2, x6, x10, zero]
);
state[3] = ReductionGate::reduce_terms(
    cs,
    [F::TWO, F::ONE, F::ONE, F::ZERO],
    [x3, x7, x11, zero]
);
    \end{lstlisting}
    \caption{Boojum version.}
    \label{fig:boojum-hclc-full}
  \end{subfigure}
  \caption{Hard-coded linear combination gate.}
  \label{fig:hard-coded-lc}
\end{figure}

\subsection{Poseidon2: hard-coded linear-combination}
Although on the surface-level the two snippets from Figure~\ref{fig:hard-coded-lc} look similar, the Clap code can compute a linear combination of an arbitrary number of terms, while the Boojum one takes as fixed number of terms (in this case 4, as the \code{ReductionGate} is instantiated with width of~4). This is due to the fact that in the Boojum case, the code is directly adding \code{ReductionGate}s (for linear combinations), which need to have a fixed size.

In the case of \clap, the \code{reduce\_terms} function is just calling the \code{add} function repeatedly, so there is no need to fix the linear combination gate size at this stage. Instead, the optimizer can detect a chain of additions, and replace it by a linear combination gate. Then, the specific instantiation of gates (the width of linear combinations in this case) can be decoupled from the circuit building.

\begin{figure*}[ht]
  \centering
  \begin{subfigure}[t]{0.40\textwidth}
    \centering
    \begin{lstlisting}[language=Rust]
fn xor(&mut self,
       l: Repr<F, bool>,
       r: Repr<F, bool>)
       -> Repr<F, bool> {
  // (l /\ ~r) \/ (~l /\ r)
  let nr = self.not(r);
  let lnr = self.and(l, nr);
  let nl = self.not(l);
  let nlr = self.and(nl, r);
  self.or(lnr, nlr)
}
    \end{lstlisting}
    \caption{\clap version.}
    \label{fig:clap-xor-full}
  \end{subfigure}
  \hfill
  \begin{subfigure}[t]{0.58\textwidth}
    \centering
    \begin{lstlisting}[language=Rust]
pub fn xor(self, cs: &mut CS, other: Self) -> Self {
    // c = ~(a /\ b) /\ ~(~a /\ ~b)
    // (1 - (a * b)) * (1 - ((1 - a) * (1 - b))) = c
    // (1 - ab) * (1 - (1 - a - b + ab)) = c
    // (1 - ab) * (a + b - ab) = c
    // a + b - ab - (a^2)b - (b^2)a + (a^2)(b^2) = c
    // a + b - ab - ab - ab + ab = c
    // a + b - 2ab = c
    // -2a * b = c - a - b
    // 2a * b = a + b - c
    let one_variable = cs.allocate_constant(F::ONE);
    let a_plus_b = compute_fma(
        cs,
        F::ONE,
        (self.variable, one_variable),
        F::ONE,
        other.variable,
    );
    let mut minus_two = F::TWO;
    minus_two.negate();
    let c = compute_fma(
        cs,
        minus_two,
        (self.variable, other.variable),
        F::ONE,
        a_plus_b,
    );
    Boolean::from_variable_checked(cs, c)
}
    \end{lstlisting}
    \caption{Boojum version.}
    \label{fig:boojum-xor-full}
  \end{subfigure}
  \caption{Full implementation of XOR.}
  \label{fig:xor-full}
\end{figure*}

\subsection{\code{XOR} Definition}
There are several ways to express the \code{xor} operation. Figure~\ref{fig:boojum-xor-full} presents the full implementations already discussed in Section~\ref{sec:experiments}.  On the left, we implement in \clap one of the simple definitions based on \code{not}, \code{and} and \code{xor}. On the right, we can see that the Boojum implementation also starts with a simple identity based on conjunction and negation, which is then elaborated in a sequence of comments until reaching an equivalent form that is friendly with the FMA constraint.

Automatic optimization (arithmetic and boolean inlining) allows us to write a simple definition for \code{xor}, while producing the same final circuit (2 FMA + 1 constant constraint). Boojum has one redundant boolean check at the end, which would be removed by common-subexpression-elimination.

Boojum also has the primitive boolean operations (\code{not}, \code{and}, \code{or}) defined too. However, the naive implementation would produce 6 FMA constraints instead of 2.

\begin{figure*}[ht]
  \centering
  \begin{subfigure}[t]{0.40\textwidth}
    \centering
    \begin{lstlisting}[language=Rust]
let mut le_4bit_chunks = [U4Var::default(); 64];
for (_idx, ((dst, src), res)) in state
    .iter_mut()
    .zip([a, b, c, d, e, f, g, h].into_iter())
    .zip(le_4bit_chunks.array_chunks_mut::<8>())
    .enumerate()
{
    let tmp = cs.add(SVar::from(*dst), src.into());
    let (_u32_tmp, chunks) =
      range_check_36_bits(cs, tmp);
    *res = chunks[..8].try_into().unwrap();
}
    \end{lstlisting}
    \caption{\clap version.}
    \label{fig:clap-sha2-full}
  \end{subfigure}
  \hfill
  \begin{subfigure}[t]{0.58\textwidth}
    \centering
    \begin{lstlisting}[language=Rust]
let mut yet_unchecked_chunks = ArrayVec::<Variable, 8>::new();
for (idx, (dst, src)) in state
    .iter_mut()
    .zip([a, b, c, d, e, f, g, h].into_iter())
    .enumerate()
{
    let tmp = compute_fma(cs, F::ONE, (one, *dst), F::ONE, src);
    let (tmp, high) = split_36_bits_unchecked(cs, tmp);
    yet_unchecked_chunks.push(high);
    if idx == 3 {
        final_d_decomposition = range_check_uint32(cs, tmp);
    }
    if idx == 7 {
        final_h_decomposition = range_check_uint32(cs, tmp);
    }
    // other variables are range-checked in next round unless
    // requested otherwise
    *dst = tmp;
}

for chunk in yet_unconstrained_chunks.chunks(3) {
    let a = chunk.get(0).copied().unwrap_or(zero);
    let b = chunk.get(1).copied().unwrap_or(zero);
    let c = chunk.get(2).copied().unwrap_or(zero);
    let _ = tri_xor_many(cs, &[a], &[b], &[c]);
}
    \end{lstlisting}
    \caption{Boojum version.}
    \label{fig:boojum-sha2-full}
  \end{subfigure}
  \caption{Range-checks in SHA2-256.}
  \label{fig:range-checks-sha2}
\end{figure*}

\subsection{SHA2-256: redundant range-check elimination and unsound oracles}
As explained in Section~\ref{sec:experiments} for Figure~\ref{fig:boojum-sha-unsound}, Boojum code uses an unsafe function (\code{split\_36\_bits\_unchecked}) to avoid having duplicated range-checks. This same pattern of using an unsound oracle and collecting unchecked variables to be checked later is also present later on in the SHA2-256 implementation, shown in Figure~\ref{fig:range-checks-sha2}.

\point{Find the bug}
The Boojum code shown in Figure~\ref{fig:boojum-sha2-full} actually contains a bug. We invite the reader to look for it.
As anticipated in Section~\ref{sec:motivation}, the error lies in the fact that the variables in the collection \code{yet\_unchecked\_chunks} are never checked. This is the result of copying-and-pasting the checks that followed the snippet from Figure~\ref{fig:boojum-sha-unsound}, which is in the same function. In this other part of the code, the name of the collection for the unchecked variables was \code{yet\_unconstrained\_chunks}, which is now checked for a second time by mistake. This means that 8 variables are not range-checked to be in the 4-bit range, meaning that a malicious prover could find different values for them affecting the result of the SHA2-256 function.

This bug has been fixed in Boojum on the following \href{https://github.com/matter-labs/era-boojum/commit/cd631c9a1d61ec21d7bd22eb74949d43ecfad0fd}{commit}, and the corrected version is already deployed in production for ZKsync Era.

\end{appendices}
\end{document}